\documentstyle[12pt,epsfig]{article}
\def\beq{\begin{equation}}
\def\eeq{\end{equation}}
\def\bea{\begin{eqnarray}}
\def\eea{\end{eqnarray}}

\def\del{\partial}
\def\eq#1{(\ref{#1})}

\def\a{\alpha}

\def\th{\theta}
\def\r{\rho}

\def\t{\tau}
\def\w{\omega}

\def\beq{\begin{equation}}
\def\eeq{\end{equation}}
\def\ber{\begin{eqnarray}}
\def\eer{\end{eqnarray}}
\def\bea{\begin{eqnarray}}
\def\eea{\end{eqnarray}}
\def\del{\partial}

\def\a{\alpha}
\def\b{\beta}
\def\g{\gamma}
\def\D{\Delta}

\def\t{\tau}

\def\r{\rho}
\def\t{\theta}
\def\w{\omega}

\def\D{\Delta}

\def\del{{\partial}}

\newfont{\Bbb}{msbm10 scaled 1200}     
\newcommand{\mathbb}[1]{\mbox{\Bbb #1}}

\def\bra{\langle}
\def\ket{\rangle}

\def\lbldef#1#2{\expandafter\gdef\csname #1\endcsname {#2}}

\def\href#1#2{#2}

\newcommand{\beqar}{\begin{eqnarray}}

\newcommand{\eeqar}{\end{eqnarray}}
\newcommand{\th}{\theta}

\def\del{\partial}

\def\a{\alpha}
\def\b{\beta}
\def\D{\Delta}

\def\t{\tau}

\def\r{\rho}
\def\th{\theta}
\def\w{\omega}

\begin{document}

\baselineskip=15.5pt
\pagestyle{plain}
\setcounter{page}{1}
\begin{titlepage}

\leftline{\tt hep-th/0112188}

\vskip -.8cm

\rightline{\small{\tt CALT-68-2363}}
\rightline{\small{\tt CITUSC/01-049}}

\begin{center}

\vskip 2 cm

{\LARGE {Boundary States for $AdS_2$ Branes in $AdS_3$}}

\vskip 2cm
{\large Peter Lee, Hirosi Ooguri, and Jongwon Park}

\vskip 1.2cm

California Institute of Technology 452-48,
Pasadena, CA 91125

\vskip 1.5cm

{\bf Abstract}
\end{center}

\noindent
We construct boundary states for the $AdS_2$
D-branes in $AdS_3$. We show that, in the semi-classical
limit, the boundary states correctly reproduce 
geometric configurations of these branes.
We use the boundary states to compute the one loop
free energy of open string stretched between the
branes. The result agrees precisely with 
the open string computation in hep-th/0106129.

\end{titlepage}

\newpage


\section{Introduction}

Recently much progress has been made in understanding
string theory in $AdS_3$, three-dimensional anti-de Sitter
space. With a background of NS-NS $B$-field, the worldsheet
is described by the $SL(2,R)$ WZW model. The spectrum of the
WZW model was proposed in \cite{first}, and it was verified
by exact computation of the one loop free energy in \cite{second}.
This has lead a proof of the no ghost theorem for string in $AdS_3$,
which had been an outstanding problem for more than 10
years.\footnote{For a list of historical
references, see the bibliography in \cite{first}.}
The unitary string spectrum emerged from these results
agrees with expectations from
the $AdS$/CFT correspondence \cite{malda, review} applied to
the case of $AdS_3$ \cite{strmalda,gks,sw}.
Correlation functions of vertex operators on the string worldsheet
were derived in \cite{teschnerone, teschnertwo, zam}. In \cite{third}, 
they are
used to compute target space correlation functions of the string
theory and the results were compared with what we expect for the
conformal field theory on the boundary of $AdS_3$. 
The correlation functions turned out to
have various singularities, and physical interpretations
were given for all of them. It was also shown that the four point
functions of the target space conformal field theory obey the
factorization rules when they should.

In this paper, we study D-branes whose worldvolume fill $AdS_2$
subspaces of $AdS_3$ \cite{stan,bachas}. We call such D-branes
as $AdS_2$ branes. From the point of view
of the CFT dual, these configurations are supposed to describe
conformal field theories in two dimensions separated by domain
walls, each of which preserves at least one Virasoro algebra \cite{bddo}.
The spectrum of open strings ending on the $AdS_2$ brane
was studied in \cite{lopt}.\footnote{
A related work was also done in \cite{petro}.} 
It was shown that, when the brane carries no fundamental string
charge, the open string
spectrum is exactly equal to the holomorphic square root of
the spectrum of closed strings in $AdS_3$ derived in \cite{first,second}. 
It contains short and long strings, and
is invariant under spectral flow. 

A boundary state \cite{ishibashi,cardy} is a useful concept
in studying D-branes. Although the construction in \cite{cardy}
assumes rationality of conformal field theory, it was shown
in \cite{zamtwo,zamthree,teschnerthree,teschnerfour} that the idea
can be applied to the non-rational case of the Liouville field
theory. More recently the technique developed in the Liouville theory
is applied to the $AdS_2$ branes \cite{chicagoone,
chicagotwo,Rajaraman,Sugawara}. 
In this paper, we will closely follow the construction 
in \cite{chicagoone,chicagotwo}. We will, however, make a different ansatz
about one point functions of closed string operators in a disk
worldsheet, which
leads to different expressions for the boundary states. 
We show that, in the semi-classical limit, these new boundary states 
reproduce the geometric configurations of the $AdS_2$ branes.

Given a pair of boundary states, it is straightforward to compute
a partition function on an annulus worldsheet. This gives the
spectrum of the conformal field theory on a strip where the
boundary conditions on the two sides of the strip are specified by
the choice of the boundary states. We can
also compute an annulus partition function when the target space
Euclidean time is periodically identified.
An integral of this partition function over
the moduli space of worldsheet then gives the one loop free energy
at finite temperature. From this, we can read off the physical
spectrum of the open string on the $AdS_2$ brane. We find that the
result agrees with the exact open string
computation in \cite{lopt}.

This paper is organized as follows. In section 2, we will compute
one point functions of closed string operators on a disk with
the boundary conditions corresponding to the $AdS_2$ branes. 
Following \cite{chicagoone,chicagotwo} closely but using 
a different ansatz, we will derive functional
relations for the one point functions. The general solution to the
relations will be given. In section 3, we will choose particular
solutions which agree with the semi-classical limit of the
$AdS_2$ branes. We will use them to compute annulus amplitudes
and derive spectral densities of open string states. We will also
compute one loop free energy to identify physical states of
open strings on the $AdS_2$ branes. We will close this paper with
some discussions on future directions. In Appendix A, we perform
some integrals used in this paper. Various coordinate systems
for $AdS_3$ are summarized in Appendix B. We also list useful
formulae of the hypergeometric function and the gamma function
in Appendix C. 

\bigskip
\noindent
{\bf Note added}: Toward the completion of this manuscript, we were
informed of a related work by B. Ponsot, V. Schomerus, and 
J. Teschner. Their result of the one point functions
on the $AdS_2$ branes agrees with ours shown in (\ref{finalexpression}). 
We thank them for communicating their results prior to
the publication. 

\section{One point functions on a disk}

The boundary state can be found by computing the one point
functions of closed string operators on a disk worldsheet.
Here we will derive them for the $AdS_2$ brane, following
the approach in \cite{chicagoone,chicagotwo}.

\subsection{Review of closed string in $AdS_3$}

In this paper, we will mostly work in Euclidean $AdS_3$, which is
the three-dimensional hyperbolic space $H_3^+$ given
by one of the two branches of  
\beq \label{hyp}
  (X^0)^2- (X^1)^2 - (X^2)^2 - (X_E^3)^2 = R^2,
\eeq
embedded in ${\bf R}^{1,3}$ with the metric
\beq ds^2 = -(dX^0)^2 + (dX^1)^2 + (dX^2)^2 + (dX_E^3)^2,
\eeq
where $R$ is the curvature radius of $H_3^+$.
It is related to $AdS_3$ with the Lorentzian signature
metric by the analytic continuation
$X^3 \rightarrow X_E^3 = -i X^3$. The Euclidean $AdS_3$ 
can also be realized as a
right-coset space $SL(2,C)/SU(2)$. Accordingly, the conformal
field theory is the $SL(2,C)/SU(2)$ coset model. 
In the semi-classical approximation, which is applicable 
when the level $k \sim R^2/\alpha'$ of the
$SL(2,C)$ current algebra is large,
states in the theory are given by normalizable functions on the
target space. The space of such states is decomposed into a sum of
principal continuous representations of $SL(2,C)$ with
$j={1\over2}+is$ ($s\in {\bf R}_+$). 
It was shown in \cite{Gawe} that the exact 
Hilbert space of the coset model at finite value of $k$ 
consists of the standard representations of $SL(2,C)$ current algebra
whose lowest energy states are given by
principal continuous representations. 

Let us introduce a convenient coordinate system ($\phi$, $\gamma$,
 $\bar\gamma$) on $H_3^+$.  They are related to the embedding coordinates 
$(X^0, X^1, X^2, X_E^3)$ in (\ref{hyp}) in the following way:
\bea
  && \phi=\log\left(X^0+X_E^3\right)/R, \\ \nonumber
  && \gamma={X^2+iX^1\over X^0+X_E^3}, \\ \nonumber
  && \bar\gamma={X^2-iX^1\over X^0+X_E^3}.
\eea
A point in the coset $SL(2,C)/SU(2)$ has its representative as
a Hermitian matrix:
 \beq
   \left( \begin{array}{cc}
      \gamma\bar\gamma e^{\phi}+ e^{-\phi} & -\gamma e^{\phi} \\
      -\bar\gamma e^{\phi} & e^{\phi} \end{array} \right),
 \eeq
and the metric can be written as
\beq
 ds^2=R^2\left(d\phi^2+e^{2\phi}d\gamma d\bar\gamma\right).
\eeq
In this theory, there exists an important set of primary fields
defined by
 \ber
\label{primary} \Phi_j(x , \bar x; z,\bar z)  = {1-2j \over \pi} (
e^{-\phi} + |\gamma - x|^2 e^\phi )^{-2j}. \eer The labels $x,\bar
x$ are introduced to keep track of the $SL(2,C)$ quantum
numbers.\footnote{In the string theory interpretation discussed in
section 3, $(x,\bar{x})$ is identified as the location of the
operator in the dual CFT on $S^2$ on the boundary of $H_3^+$.}
The $SL(2,C)$ currents act on it as
 \beq
 \label{current_action}
   J^a(z)\Phi_j(x,\bar x;w,\bar w) \sim -\frac{D^a}{z-w}\Phi_j(x,\bar x;w,\bar w)\,,\qquad a=\pm\,,3\,,
 \eeq
where $D^a$ are differential operators with respect to $x$ defined
as
 \beq
  D^+ =\frac{\del}{\del x},\qquad D^3 = x\frac{\del}{\del x}+j,\qquad D^- =  x^2\frac{\del}{\del
  x}+2jx.
 \eeq
The energy momentum tensor is given by the Sugawara construction,
and the worldsheet conformal weights of this operator is \beq
\label{weight} \Delta_j = -{ j(j-1)\over k-2}. \eeq 
When $j={1 \over 2} + is$ with $s \in {\bf R}$,
the operators (\ref{primary}) correspond to the normalizable
states in the coset model. 
The operators with the $SL(2,C)$ spin $j$ and $(1-j)$ are
not independent but are related to each other by the following reflection
symmetry relation,
\beq \label{reflection}
\Phi_j(x,\bar x;z,\bar z) = {\mathcal
R}(j)\frac{2j-1}{\pi}\int d^2x^\prime |x-x^\prime |^{-4j}
\Phi_{1-j}(x^\prime,\bar x^\prime;z,\bar z), \eeq where
 \beq
   {\cal R}(j)=\nu^{1-2j} {\Gamma\left(1-{2j-1\over k-2}\right)\over
    \Gamma\left(1+{2j-1\over k-2}\right)}.
 \eeq

The two and three point functions of these operators have been
computed in \cite{teschnerone, teschnertwo, zam}. The two point
function has the form, \ber \label{reflectionrelation}
&& \langle \Phi_j(x_1,\bar
x_1;z_1,\bar z_1 )
 \Phi_{j'}(x_2,\bar x_2;z_2,\bar z_2 ) \rangle\nonumber \\
&& = { 1\over |z_{12}|^{4\Delta_j}} \left[ \delta^2(x_1-x_2)
\delta(j+j'-1) +{B(j)\over |x_{12}|^{ 4j }} \delta(j-j')\right].
\eer The coefficient $B(j)$ is given by \ber B(j) =  {k-2 \over
\pi} { \nu^{1-2j} \over \gamma\left( { {2j -1\over k-2}  }\right)} ,\eer
where \ber \gamma(x) = {\Gamma(x) \over \Gamma(1-x) }.\eer
The choice of $\nu$ will not play an important
role in the discussion of this paper, except for its behavior
in the semi-classical limit
 \beq \label{classical} \nu \rightarrow 1 ~~~~(k \rightarrow \infty),\eeq 
which can be deduced by comparing (\ref{reflection})
with its classical counterpart \cite{teschnerclassical}.  

The three point function is expressed as \ber  &&\langle \Phi_{j_1}
(x_1,\bar x_1;z_1,\bar z_1 )
 \Phi_{j_2}(x_2,\bar x_2; z_2,\bar z_2)
\Phi_{j_3}(x_3,\bar x_3;z_3,\bar z_3 )
\rangle  = \nonumber\\
&&=C(j_1,j_2,j_3)
{1
 \over | z_{12}|^{2(\Delta_1+ \Delta_2 - \Delta_3) }
|z_{23}|^{2(\Delta_2+ \Delta_3 - \Delta_1) }
|z_{31}|^{2(\Delta_3+ \Delta_1 - \Delta_2)}} \times\nonumber\\
&&~~~~~~~{1 \over
|x_{12}|^{2(j_1+j_2-j_3)}|x_{23}|^{2(j_2+j_3-j_1)}
|x_{31}|^{2(j_3+j_1-j_2)} }  ,
\eer
with the coefficient $C(j_1,j_2,j_3)$ given by
\ber
 C(j_1,j_2,j_3) = -{
G(1-j_1-j_2-j_3) G(j_3-j_1-j_2) G(j_2-j_3-j_1)
G(j_1-j_2-j_3) \over   2\pi^2
 \nu^{j_1+j_2+j_3-1}
\gamma\left( {k-1\over k-2}\right)G(-1) G(1-2j_1) G(1-2j_2)G(1-2j_3)},\eer
where
\ber
 G(j) = (k-2)^{{j(k-1-j) \over 2(k-2)}} \Gamma_2(
-j ~|~1,~ k-2)
 \Gamma_2(k-1+j ~|~1,~ k-2) ,\eer
and $\Gamma_2(x| 1, \omega)$ is the Barnes double Gamma function
defined by
\ber &&
  \log \left(\Gamma_2(x|1,\omega)\right)\nonumber\\
&&= \lim_{\epsilon \rightarrow 0}
 {\partial \over \partial \epsilon}\left[
 \sum_{n,m=0}^\infty
(x+n+m\omega)^{-\epsilon}
-
 \sum_{n,m = 0 \atop  (n,m)\neq(0,0)}
(n+m\omega)^{-\epsilon}\right].\eer

\subsection{Constraints on one point functions}

In this subsection, we will derive functional equations
satisfied by one point functions of closed string operators
on a disk with boundary conditions corresponding to  $AdS_2$ 
branes. We will follow the discussion in \cite{chicagoone,
chicagotwo} closely, except that we use a more general
 ansatz for the
one point function as we will explain in the next paragraph.   
Given the one point functions, we can
find boundary states $|B\rangle\rangle$ for the $AdS_2$ branes.  
According to \cite{bachas}, $AdS_2$ branes preserve
one half of the current algebra symmetry because of the boundary
condition on the currents, 
\beq
  J^a(z)=\bar J^a(\bar z); \;\;\;\;
  \mbox{at  } z=\bar z, \;\;\;\;
  a=3,+,-.
\eeq 
In the closed string channel, it is translated into the condition
on the boundary states as\beq
 \label{gluing}
 (J^a_n +\bar J^a_{-n}) |B\rangle\rangle = 0, \;\;\;\;a=3,+,-,\;\;n\in{\bf Z}.
\eeq

We start with the ansatz that the one point function is of the form
\bea\label{ansatz}
  \langle \Phi_j(x,\bar x;z, \bar z) \rangle=
  {U^{+} (j) \over |x-\bar x|^{2j} |z-\bar z|^{2\D_j}}
  \;\;\; \mbox{for  } Im~x>0,  \\ \nonumber
  {U^{-} (j) \over |x-\bar x|^{2j} |z-\bar z|^{2\D_j}}
  \;\;\; \mbox{for  } Im~x<0.
\eea The $z$ dependence is determined from conformal invariance 
on the worldsheet and the $x$ dependence is fixed by 
conformal invariance on the target space. The parameters 
$(x, \bar x)$ can be regarded as
coordinates on the boundary of
Euclidean $AdS_3$, which is $S^2$.  The $AdS_2$ brane divides $S^2$ into
half, and the upper half plane covers one patch and the lower half
the other. From the point of view of the 
conformal field theory on the boundary,
the $AdS_2$ introduces an one-dimensional defect on $S^2$, across
which the two different CFT's are glued together \cite{bddo}.
Therefore the one point function $\langle \Phi_j \rangle$ may
have a discontinuity across $Im~ x= 0$. The expression
 (\ref{ansatz}) allows such a discontinuity since the coefficients
$U^+$ and $U^-$ can be different. 

The reflection symmetry (\ref{reflection}) implies
a relation between $U^+$ and $U^-$. Taking the expectation
value of (\ref{reflection}) on both sides, we get
 \bea
  \label{oneptreflect}
  {U^{\pm}(j) \over |x-\bar x|^{2j}}& =& {\cal R}(j) {2j-1\over \pi}
      \left( \int_{Im~x>0} d^2x' U^+(1-j) {|x-x'|^{-4j}\over
    |x'-\bar x'|^{2(1-j)}} \,\, + \right. \\ \nonumber && \left.
    ~~~~~~~~~~~~~~~~~~~~
~~+ \int_{Im~x<0} d^2x' U^-(1-j) {|x-x'|^{-4j}\over |x'-\bar
    x'|^{2(1-j)}} \right) \,,
 \eea
On the left hand side, we choose $U^+$  for $Im~ x>0$ 
and $U^-$ for $Im~x <0$. 
Setting $x'= x'_1+i x'_2$, we can rewrite the
above $x'$-integration as
 \bea
  &=& U^+(1-j) \int_{-\infty}^\infty dx'_1 \int_0^\infty dx'_2 {\left({x'_1}^2
   +(x'_2-x_2)^2 \right)^{-2j} \over |2 x'_2|^{2(1-j)}} +
   \\ \nonumber
  && + U^-(1-j) \int_{-\infty}^\infty dx'_1 \int_{-\infty}^0 dx'_2 
  {\left({x'_1}^2+(x'_2-x_2)^2 \right)^{-2j} \over |2 x'_2|^{2(1-j)}} .
 \eea
By using the Euler integral
 \bea
   \int_0^1 dx \, x^{a-1} (1-x)^{b-1}={\Gamma(a) \Gamma(b) \over \Gamma(a+b)},
 \eea
we see that for the case $x_2>0$, the first term vanishes and
 only the second term
contributes.  On the other hand, when $x_2<0$, the second term
vanishes and only the first term contributes.  We can summarize
the result as
 \beq
 \label{refsym_U}
   U^{\pm}(j)={\cal R}(j) U^{\mp}(1-j).
 \eeq
Introducing $f^\pm(j)$ by
 \beq
  U^\pm(j)=\Gamma\left(1-{2j-1\over k-2}\right)
\nu^{{1\over 2}-j}f^\pm(j)\, ,
 \eeq
the reflection relation (\ref{refsym_U}) becomes
 \beq
 \label{refsym_f}
  f^\pm(j) = f^\mp(1-j)\,.
 \eeq

To determine $f^+(j)$, it is useful to consider the
bulk two point function of $\Phi_j$ with the degenerate field
with $j=-1/2$, which satisfies
\beq
  \del^2_x \Phi_{-{1\over 2}} = 0.
\eeq 
It implies that there are only two terms
in its operator product expansion with $\Phi_j$ as
 \beq\label{degenerateexpansion}
  \Phi_{-{1\over 2}} \Phi_j \sim C_-(j) \Phi_{j-{1\over 2}} +
    C_+(j) \Phi_{j+{1\over 2}}+ \cdots,
\eeq 
Here the ``$\cdots$
'' denote the current algebra descendants.
To simplify the equations, in this subsection, 
we are suppressing the dependence on $x$ and $z$.
The coefficients  $C_\pm(h)$ have been derived
in the earlier literature, but let us compute them here for
completeness. Let us take a correlation function of 
$\Phi_{j'}$ with both sides of (\ref{degenerateexpansion}). 
The left hand side gives
 \beq \label{three}
  \langle \Phi_{-{1\over 2}} \Phi_j \Phi_{j'}\rangle =
  C\left(-{1\over 2},j,j'\right).
\eeq
Using the expression for the three point function
and the identity 
\beq
 \lim_{\epsilon\rightarrow 0} {G(j_1-j_2+\epsilon) G(j_2-j_1+\epsilon)
 \over G(-1) G(1+2 \epsilon)}=-2\pi(k-2) \gamma\left({k-1 \over k-2}
 \right)\delta(j_1-j_2),
\eeq
we find that (\ref{three}) contains delta functions
at $j'=j \pm {1\over 2}$. 
On the other hand, the right hand side of (\ref{degenerateexpansion})
gives 
\beq
  \langle \Phi_{-{1\over 2}} \Phi_j \Phi_{j'}\rangle \sim
  \delta\left(j'-j+{1\over 2}\right) C_-\left(j\right) B\left(j-{1\over 2}\right)
 +\delta\left(j'-j-{1\over 2}\right) C_+(j) B\left(j+{1\over 2}\right)
 +\cdots .
\eeq 
Comparing the coefficients of the delta functions, we find
\bea\label{coefficienttwo}
  C_+(j)&=&\nu {\gamma\left(-{1\over k-2}\right)
\over \gamma\left(-{2 \over k-2}\right)} , \\ \nonumber
  C_-(j)&=&{\gamma\left(-{1 \over k-2}\right) 
\gamma\left({2j-2 \over k-2}\right) \over
          \gamma\left(-{2 \over k-2}\right) 
  \gamma\left({2j-1 \over k-2}\right)}.
\eea

Given the coefficient
$C_\pm(j)$ for the operator product expansion
of $\Phi_{-{1\over 2}} \Phi_j$, one can deduce
a functional relation for $U^+(j)$.\footnote{Here we take
$x$ to be on the upper half plane. A similar relation
for $U^-$ can be derived by considering $x$ in the
lower half plane. Since $U^\pm$ are related to each
other by the reflection relation, it is sufficient
to determine conditions on $U^+$.}  Consider the following
bulk two point function involving a degenerate field $\Phi_{-1/2}$
 \beq
   \langle \Phi_{-{1\over 2}}(x;z) \Phi_j (x';z') \rangle.
 \eeq
Using the operator 
product expansion derived in the above paragraph, 
we find
 \bea
  \label{twopt}
  \langle \Phi_{-{1\over 2}}(x;z) \Phi_j (x';z') \rangle &&=
  {|x'-\bar x'|^{-1-2j} \over |x-\bar x'|^{-2}}
  {|z'-\bar z'|^{-{3u\over 2}-2\D_j} \over |z-\bar z'|^{-3u}}
  \times \\ \nonumber &&
  \left(C_+(j) U^+\left(j+{1\over 2}\right) {\cal
  F}_+(x,x';z,z')+C_-(j) U^+\left(j-{1\over 2}\right) {\cal
  F}_-(x,x';z,z')\right).
 \eea
Here ${\cal F_{\pm}}$ are four-point conformal blocks.
According to \cite{teschnerone}, we have
 \bea
 {\cal F}_+(x,x';z,z')& =& \eta^{u(1-j)}(1-\eta)^{-{u\over 2}}
   \Big( \chi F(-u,1-2uj,1-u(2j-1);\eta)+ 
  \\ \nonumber && ~~~~\left. + {u\eta\over 1+u(1-2j)} 
  F(1-u,1-2uj,2-u(2j-1);\eta)\right), \\
 {\cal F}_-(x,x';z,z')&=&  \eta^{uj}(1-\eta)^{-{u\over 2}}
     \left( \chi {1\over 2j-1} F(2u(j-1),1-u,u(2j-1)+1;\eta)+
   \right. \\ \nonumber && ~~~~+ F(1-u,-u,u(2j-1);\eta) \Big), 
 \eea
where $F(a,b,c;z)$ are the hypergeometric functions
${}_2F_1(a,b,c;z)$,
\beq \label{whatu} u = {1 \over k-2},\eeq
and $\chi$ and $\eta$ are cross ratios of the target space
and the worldsheet coordinates,
 \beq
   \chi={|x-x'|^2\over |x-\bar x'|^2}, \;\;\;\;
   \eta={|z-z'|^2\over |z-\bar z'|^2}.
 \eeq

It was pointed out in \cite{chicagoone} that, when
the operator $\Phi_{-{1\over 2}}$ approaches the boundary
of the worldsheet, it overlaps
only with the identity operator and the operator with 
$j=-1$. This is analogous to the situation in the Liouville
theory discussed in \cite{zam}.
Using this, the two point function $\langle \Phi_{-{1\over 2}}
\Phi_j\rangle$ can be evaluated as
 \bea
  \label{twoptcross}
  \langle \Phi_{-{1\over 2}}(x;z) \Phi_j (x';z') \rangle &= &
  {|x'-\bar x'|^{-1-2j} \over |x-\bar x'|^{-2}}
  {|z'-\bar z'|^{-{3u\over 2}-2\D_j} \over |z-\bar z'|^{-3u}}
  \times \\ \nonumber &&
  \left( B_+(j) {\cal G}_+(x,x';z,z') + B_-(j) 
   {\cal G}_-(x,x';z,z')\right),
 \eea
where
 \bea
 {\cal G}_+(x,x';z,z')&=&  \eta^{uj}(1-\eta)^{3u\over 2}
     \Big( (1-\chi) F(1+u,2uj,1+2u;1-\eta)-
    \\ \nonumber && \left. (1-\eta){2uj \over 1+2u} 
   F(1+u,1+2uj,2(1+u);1-\eta) \right), \\
 {\cal G}_-(x,x';z,z')& =& \eta^{uj}(1-\eta)^{-{u\over 2}}
   \left( {1-\chi \over 2} F(2u(j-1),1-u,1-2u;1-\eta)+ \right.
  \\ \nonumber &&  F(2u(j-1),-u,-2u;1-\eta)\Big),
 \eea
and $u$ is given by (\ref{whatu}).
The conformal blocks in the two expressions,
(\ref{twopt}) and (\ref{twoptcross}), are related to
each other as
\bea
 && {\cal F}_- = {\Gamma(u(2j-1)) \Gamma(1+2u) \over
    \Gamma(2uj) \Gamma(1+u)} {\cal G}_- +
   {\Gamma(u(2j-1))\Gamma(-2u)\over \Gamma(2u(j-1))
    \Gamma(-u)} {\cal G}_+, \\
 && {\cal F}_+ = {\Gamma(1+u(1-2j)) \Gamma(1+2u)\over
     \Gamma(1+2u(1-j)) \Gamma(1+u)} {\cal G}_- -
   {\Gamma(1+u(1-2j)) \Gamma(-2u)\over \Gamma(1-2uj)
    \Gamma(-u)} {\cal G}_+.
\eea
(See appendix \ref{hypergeometric} for 
some useful identities involving the hypergeometric function.)
According to \cite{zamtwo,chicagoone,chicagotwo}, $B_+$ is given by
\beq
 B_+(j)=A_0 U^+(j)\,,
\eeq 
where $A_0$ is a constant that depends on the boundary
condition and can be interpreted as the fusion coefficient of
$\Phi_{-1/2}$ with the boundary unit operator.  
Likewise, $B_-(j)$ term can be interpreted as the contribution
coming from the fusion with $j=-1$ boundary operator.  Comparing the
terms dependent on ${\cal G}_+$ in (\ref{twopt}) and
(\ref{twoptcross}),
 we derive the following 
functional equation for $U^+(j)$:
 \bea
   A_0 U^+(j)&=&{\Gamma(-2u)\over \Gamma(-u)}
  \left(C_-(j)U^+\left(j-{1\over 2}\right){\Gamma(u(2j-1))
  \over \Gamma(2u(j-1))} \,\,- \right. \\ \nonumber && \left.
  C_+(j)U^+\left(j+{1\over 2}\right)
  {\Gamma(1+u(1-2j))\over \Gamma(1-2uj)}\right).
 \eea
Substituting the expression for $C_\pm(j)$ given by (\ref{coefficienttwo}),
this reduces to
 \beq
  \label{funct1}
  A_0 \nu^{-1/2}{\Gamma\left(1 + {1 \over k-2}\right)
\over \Gamma\left(1+{2\over k-2}\right)}
f^+(j)=f^+\left(j-{1\over 2}\right)-f^+\left(j+{1\over
  2}\right) .
 \eeq
Given $f^+(j)$ satisfying (\ref{funct1}), 
we get $f^-(j)$ by using (\ref{refsym_f}).

It is convenient to introduce a parameter $\Theta$ by
\beq \label{whattheta}
A_0 \nu^{-1/2} {\Gamma\left(1 + {1 \over k-2}\right)
\over \Gamma\left(1+{2\over k-2}\right) }
= -2\sinh \Theta.
\eeq
We can regard $\Theta$ as parametrizing the boundary
condition as $A_0$ depends on it. In fact, the semi-classical analysis in
the next section shows that $\Theta$ specifies
 the location of the $AdS_2$ brane. 
A general solution to (\ref{funct1}) and (\ref{refsym_f})  is
then a linear combination of 
 \bea \label{solutions}
  f^{\pm}(j) = \exp\left[\pm(\Theta+i 2n\pi )(2j-1)\right]~,~~~
 \exp\left[\pm(i\pi(2n+1)-\Theta)(2j-1)\right],
\eea
 where  $n\in {\bf Z}$.

\section{Boundary states for $AdS_2$ branes}

Among the solutions (\ref{solutions}), we claim that
the one which correctly represents the boundary
condition for a single $AdS_2$ brane is
 \beq
  \label{oursolution} f^\pm_\Theta(j) = Ce^{\pm\Theta(2j-1)}, \eeq
 where
 $C$ is a constant
independent of $j$ but may depend on $\Theta$ and $k$. From now
on, we neglect this constant and it will not affect the rest of 
our discussion. In this section, we will provide evidences for
this claim.

Given the $f^\pm(j)$ solution (\ref{oursolution}), 
and therefore the one point function $U^\pm(j)$, 
the boundary state is  expressed as
 \beq \label{bs}
   |B\rangle\rangle_\Theta = \int_{{1\over 2}+i {\bf R}^+} dj
    \left(  \int_{Imx>0} d^2x {U^+_\Theta(1-j)\over |x-\bar x|^{2(1-j)}}
   |j,x,\bar x\rangle\rangle_I +
              \int_{Imx<0} d^2x {U^-_\Theta(1-j)\over |x-\bar x|^{2(1-j)}}
   |j,x,\bar x\rangle\rangle_I \right),
 \eeq
where $|j,x,\bar x\rangle\rangle_I$ is an Ishibashi state built
 on the primary state $|j,x,\bar x\rangle$.  The coefficients
are chosen so that the one point functions are correctly reproduced
as
\beq
  \langle j,x,\bar x| B \rangle\rangle_\Theta =
  \lim_{z \rightarrow \infty} |z-\bar z|^{2\Delta_j}
\langle \Phi_j(x, \bar x; z, \bar z) \rangle.
\eeq 
It can be verified using the two point function
(\ref{reflectionrelation}) and the reflection relation
(\ref{refsym_U}) that the boundary state given by
(\ref{bs}) satisfies this condition. In the following,
we will examine aspects
of $|B \rangle\rangle_\Theta$.

\subsection{Semi-classical analysis}

Let us first study the semi-classical limit of the
boundary state for the solution (\ref{oursolution}).  
We use the method of \cite{mms1,chicagotwo} to identify
the D-brane configuration corresponding to the boundary
state.  In the semi-classical limit
($k\rightarrow\infty$), we can consider a closed string state
$|g\ket$  localized at $g \in H^+_3$. Namely it is defined
so that 
\beq \label{localized}
 \bra g|j,x,\bar x\ket = \Phi_j(x,\bar x|g).
 \eeq
The overlap of the localized state $|g\ket$ with the boundary
state $|B\rangle\rangle_\Theta$ can then tell us about the
configuration of the brane. (An analogous idea has been used in \cite{ooy}
to characterize boundary states for D-branes wrapping 
on cycles in Calabi-Yau manifolds.)
In the limit $k \rightarrow \infty$,
the overlap $\langle g | B\rangle\rangle_\Theta$ is simplified
as
\begin{eqnarray*}
&&\lim_{k\rightarrow \infty} \bra g|B\ket\ket_\Theta \\
&&=  \int_{{1\over2}+i{\mathbf R}^+}dj\left(\int_{Im x>0}d^2x
\frac{f^+_{\Theta}(1-j)}{|x-\bar x|^{2(1-j)}}\Phi_j(x,\bar x|g) +
\int_{Im x<0}d^2x \frac{f^-_{\Theta}(1-j)}{|x-\bar
x|^{2(1-j)}}\Phi_j(x,\bar x|g) \right),
\end{eqnarray*}
where we used (\ref{classical}). 
First let us focus on the integral on the upper half plane,
\beq \int_{{1\over2}+i{\mathbf R}^+}dj \frac{2j-1}{\pi}
\left(\int_{x_2>0}d^2x
\frac{f^+_{\Theta}(1-j)}{(2x_2)^{2(1-j)}}\frac{1}{\left[
e^\phi(x_1^2 +x_2^2-(\gamma+\bar\gamma)x_1+i(\gamma-\bar\gamma)x_2
+\gamma\bar\gamma)+e^{-\phi}\right]^{2j}} \right) \nonumber \,.
\eeq 
After integrating over $x_1$ and changing integration
variable $x_2^\prime = e^\phi x_2$, we get 
\beq
 \int_{{1\over2}+i{\mathbf R}^+}dj
\frac{2j-1}{\sqrt{\pi}2^{2-2j}}\frac{\Gamma(2j-{1\over
2})}{\Gamma(2j)}f^+_{\Theta}(1-j) \int_0^\infty dx_2^\prime
(x_2^\prime)^{2j-2}\left[\left(x_2^\prime +
{i\over2}(\gamma-\bar\gamma)e^\phi\right)^2+1\right]^{{1\over2}-2j}
\nonumber \,. 
\eeq 
In terms of $AdS_2$ coordinates defined in Appendix B, we have
\beq
  {i \over 2}(\gamma -\bar \gamma) e^\phi = -\sinh\psi.
\eeq 
The $x_2^\prime$-integral is performed in appendix A, and the result is 
\beq \int_{{1\over2}+i{\mathbf R}^+}dj f^+_{\Theta}(1-j)
\frac{e^{\psi(2j-1)}}{\cosh\psi} = \int_0^\infty ds
\frac{e^{2i(\psi-\Theta)s}}{\cosh\psi}, 
\eeq  
where $j=1/2+is$.  Likewise, we can
perform the integral over the lower half plane. Combining the results
together, we find that the overlap is given by
 \bea \lim_{k\rightarrow \infty} \bra g|B\ket\ket_\Theta &=&
\int_0^\infty ds \frac{e^{2i(\psi-\Theta)s}}{\cosh\psi} +
\int_0^\infty ds \frac{e^{-2i(\psi-\Theta)s}}{\cosh\psi} =
\int_{-\infty}^{\infty} ds \frac{e^{2i(\psi-\Theta)s}}{\cosh\psi}\\
&=& \frac{1}{4\pi \cosh\psi}\delta(\psi-\Theta) 
= {1\over 4\pi} \delta(\sinh\psi
- \sinh\Theta) . \eea
Thus we found that $\langle g | B \rangle\rangle_\Theta$ 
has a 
support in the two-dimensional subspace at $\psi = \Theta$ in 
$AdS_3$. Namely the parameter $\Theta$ of the boundary state
specifies the location of the $AdS_2$ brane.

In this formalism, the insertion of the identity operator
should reproduce the Born-Infeld action in the semi-classical regime.  
Using the reflection symmetry and taking the large $k$ limit in
(\ref{oneptreflect}), we see that
 \beq
    \langle 1 \rangle_\Theta \propto
    \cosh\Theta\int d^2x |x-\bar x|^{-2}.
 \eeq
The Born-Infeld action of $AdS_2$ branes has been computed by
independent methods in \cite{bachas} to be
 \beq
   S_{BI} \propto \cosh\psi \int d\omega dt \cosh\omega.
 \eeq
If we interpret $\int d^2x |x-\bar x|^{-2}$ as the volume
of $AdS_2$ branes, the two expressions agree with the
identification $\psi =\Theta$.

The one point function is given by a linear combination of
the solutions (\ref{solutions}) to the functional equations
discussed in the last section. Since we found that
 $f_\Theta^\pm \sim  e^{\pm\Theta(2j-1)}$
reproduces the correct semi-classical geometry of
the $AdS_2$ brane,
the coefficients for all other solutions in (\ref{solutions})
should vanish in the semi-classical limit $k \rightarrow \infty$. 
In fact we can make a stronger statement. If we assume the state
$|g \rangle$ satisfying (\ref{localized}) exists at finite value 
of $k$, then the $j$-integral to compute overlap 
$\langle g | B \rangle\rangle_\Theta$ is finite only for the 
particular solution, 
$f_\Theta^\pm(j)$. For all other solutions in
(\ref{solutions}), the integral is divergent for $k>3$. 
This suggests that the coefficients in front of the other
solutions in (\ref{solutions}) should vanish identically
even for finite $k$.

\subsection{Annulus amplitudes}

Given the boundary states, the partition function on the annulus
worldsheet with the boundary conditions $\Theta_1$ and $\Theta_2$
on the two boundary of the annulus is computed as exchanges
of closed string states between $|B\rangle\rangle_{\Theta_1}$
and $|B\rangle\rangle_{\Theta_2}$. If we view this in
the open string channel, one should be able to express it
as a sum over states of open string stretched between
the two $AdS_2$ branes at $\Theta_1$ and $\Theta_2$. These
open string states are normalizable. For the Euclidean $AdS_2$
branes in Euclidean $AdS_3$, they belong to principal continuous
representations. 

We want to compute the following amplitude between two boundary
states: \bea \nonumber & &
{}_{\Theta_1}\bra\bra B|(q_c^{1\over2})^{L_0+\bar L_0 -{c\over12}}
|B\ket\ket_{\Theta_2}\\ \nonumber &=& {\int}_{{1\over2}+i{\bf
R}^+}dj
 \left( \int_{Im x>0} d^2x \frac{ {U^+}_{\Theta_1}(j) {U^+}_{\Theta_2}(1-j)}{ |x-\bar x|^2 }
+ \int_{Im x<0}d^2x \frac{ {U^-}_{\Theta_1}(j) {U^-}_{\Theta_2}(1-j)}{|x-\bar x|^2}
\right) \frac{q_c^{\frac{s^2}{k-2}}}{\eta(q_c)^3} \\
\nonumber &=& \int d^2 x |x-\bar x|^{-2} \cdot
\int_0^\infty ds
\frac{\cos2(\Theta_1-\Theta_2)s}{\sinh\frac{2\pi}{k-2}s}
\frac{2\pi}{k-2}s \frac{q_c^{\frac{s^2}{k-2}}}{\eta(q_c)^3}. \eea
Using the fact \beq \nonumber s
\frac{q_c^{\frac{s^2}{k-2}}}{\eta(q_c)^3} =
\frac{2\sqrt{2}}{\sqrt{k-2}}\int_0^\infty ds^\prime
\sin\left(\frac{4\pi}{k-2}ss^\prime\right)s^\prime
\frac{q_o^{\frac{{s^\prime}^2}{k-2}}}{\eta(q_o)^3}\,\,, \eeq where
\beq q_c = e^{2\pi i \tau_c}, \qquad q_o = e^{2\pi i \tau_o},
\qquad \tau_o = -\frac{1}{ \tau_c},
\eeq
we can go to the open string channel:
\bea
\nonumber
&=&  \int d^2 x |x-\bar x|^{-2}\cdot \frac{4\sqrt{2}\pi}{(k-2)^{3/2}}\int_0^\infty ds^\prime \left(\int_0^\infty ds \frac{\cos\left(2\Theta_{12}s\right)\sin\left(\frac{4\pi}{k-2}ss^\prime\right)}{\sinh\frac{2\pi}{k-2}s}\right) s^\prime\frac{q_o^{\frac{{s^\prime}^2}{k-2}}}{\eta(q_o)^3} \\
\nonumber &=& \int d^2 x |x-\bar x|^{-2}\cdot
\frac{\pi}{\sqrt{2}\sqrt{k-2}} \int_0^\infty ds^\prime
\frac{2s^\prime\sinh2\pi
s^\prime}{ \cosh\left((k-2)\Theta_{12}\right)+\cosh\left(2\pi s^\prime\right)}
\frac{q_o^{\frac{{s^\prime}^2}{k-2}}}{\eta(q_o)^3},
\eea
where $\Theta_{12} = \Theta_1-\Theta_2$.  Obviously, the overall factor
involving the $x$-integral is divergent. We can identify
it as the volume divergence coming
from the fact that the $AdS_2$ brane is non-compact.
This divergence also reflects the fact that normalizable states
in the open string sector belong to principal continuous representations
which are infinite dimensional.\footnote{Here we 
have focused on the part of
the annulus amplitude that scales as the volume of the $AdS_2$
brane. After the completion of the manuscript, we were
informed by B. Ponsot, V. Schomerus, and J. Teschner
that, under a certain regularization of the volume divergence, 
one finds a finite additive term to the annulus partition function
with nontrivial dependence on $s$ and $\Theta$.
This would add corrections to the spectral density (\ref{spectraldensity})
that do not scale as the volume of the $AdS_2$ brane.}

If we define $\rho(s)$ as the density of 
states in open string Hilbert space belonging to the
current algebra representation whose ground states are in
principal continuous representation with $j={1\over 2} + is$,
the above result shows that it is given by
\beq \label{spectraldensity}
\rho(s) \propto \frac{s \sinh2\pi
s}{ \cosh\left((k-2)\Theta_{12}\right)+\cosh\left(2\pi s\right)}\, .
\eeq
 The spectral density is real and
non-negative as it should be. For the case $\Theta_1 = \Theta_2$, 
$\Theta_1$ dependence completely
disappears.
Note, however, we have neglected the overall constant $C$ in
the one point function (\ref{oursolution}), which can be
$\Theta$-dependent but is independent of $s$.


\subsection{One loop free energy at finite temperature}
In this subsection, we consider $AdS_2$ branes in
finite-temperature $AdS_3$ and compute the partition function by
using the boundary states we have constructed. For the boundary
state with $\Theta=0$, 
we can directly compare the result with that of appendix A
in \cite{lopt}. 

Finite-temperature $AdS_3$ is given by identifying the Euclidean
 time $t_E= it$
in the target space as,
 \beq \label{identi} t_E \sim t_E +\beta\,. \eeq It induces
 identification of boundary coordinates as well, \beq
 \label{identification}
|x| \sim
|x|e^{\beta}\,. \eeq 
The thermal
identification introduces new sectors of closed strings winding around the
compact time direction \cite{Rajaraman}. As shown in \cite{first}, 
the winding sectors are generated by the spectral flow
  \beq
    g \rightarrow e^{{i\over 2} w_R x^+ \sigma_2} g
    e^{{i\over 2} w_L x^- \sigma_2},
  \eeq 
where $\sigma_1, \sigma_2$, and $\sigma_3$ are the Pauli matrices.  
If we choose $w_R=-w_L = -i w \beta/2\pi$ with $w\in {\bf Z}$,
the action generates the following change in the string coordinates
in Euclidean $AdS_3$,
\bea
  t_E &\sim& t_E + {w \beta \over 2\pi} \sigma , \nonumber \\
  \theta & \sim & \theta - {w\beta \over 2\pi } \tau , \nonumber \\
  \rho & \sim & \rho.
\eea
The string worldsheet remains periodic in $\sigma \rightarrow \sigma + 2\pi$,
modulo the identification (\ref{identi}).  This induces the
following transformation on the Virasoro generator,
\bea \label{sugawaraflow}
&& L_0 \rightarrow L_0 +iw\frac{\beta}{2\pi}J^3_0
+kw^2\frac{\beta^2}{16\pi^2}, \\ \nonumber && \bar L_0 \rightarrow
\bar L_0 -iw\frac{\beta}{2\pi}\bar J^3_0
+kw^2\frac{\beta^2}{16\pi^2}. \eea

Correspondingly the boundary states include
all the winding sectors: 
\beq |B; \beta\ket\ket_\Theta = \sum_w
|B;\beta\ket\ket_{\Theta,w}.
\eeq 
Here $|B;\beta\ket\ket_{\Theta,w=0}$ is the
boundary state given by (\ref{bs}), except that
the $x$ integral is restricted in the range
\beq \label{range}
e^{-\beta} \leq |x| \leq 1,\eeq which is the fundamental domain
of the identification (\ref{identification}). The other
states $|B;\beta\ket\ket_{\Theta,w}$ are
given by performing the spectral flow (\ref{sugawaraflow}).
The amplitude we want to compute then is 
\beq
{}_{\Theta_1}\bra\bra B;\beta|(q_c^{1\over2})^{L_0+\bar L_0
-{c\over12}}|B;\beta\ket\ket_{\Theta_2} = \sum_{w=-\infty}^\infty
{}_{\Theta_1,w}\bra\bra B;\beta | (q_c^{1\over2})^{L_0+\bar L_0 -{c\over12}}
|B; \beta \ket\ket_{\Theta_2,w}.
\eeq 
The overlap in the $w$ winding sector can be expressed as
\bea \nonumber
&& {}_{\Theta_1,w}\bra\bra B;\beta|
(q_c^{1\over2})^{L_0+\bar L_0 -{c\over12}} |B;\beta\ket\ket_{\Theta_2,w}
\nonumber \\
& =& q_c^{\frac{k\beta^2}{16\pi^2}w^2} {}_{\Theta_1,w=0}\bra\bra B;\beta
|(q_c^{1\over2})^{L_0+\bar L_0 -{c\over12}}
e^{\frac{\beta\tau_c}{2}w(-J^3_0+\bar J^3_0)} |B;\beta\ket\ket_{\Theta_2,w=0},
\eea
where $\tau_c=it_c$ and $\phi_0 =- \beta t_c w/2$.  Substituting
(\ref{bs}) into this, we find that the right-hand side is expressed
as an integral over $x$ in the range (\ref{range}). The integration
domain is divided into four regions, depending on the signs of
$Im x $ and $Im (e^{i\phi_0} x)$. Combining them together, we find
\bea
 \nonumber && -2q_c^{\frac{k \beta^2}{16\pi^2}w^2}
{\int}_{{1\over2}+i{\bf R}^+}dj \left( \int_{Im x >0, \,
Im(xe^{i\phi_0})>0,\, e^{-\beta}<|x|<1} d^2 x
|x-\bar x|^{-2j}|xe^{i\phi_0}-\bar xe^{-i\phi_0}|^{-2(1-j)}\right) \times \\
\nonumber & &\,\,
\left( U^+_{\Theta_1}(j) U^+_{\Theta_2}(1-j) + U^+_{\Theta_1}(j)
U^-_{\Theta_2}(1-j) + U^-_{\Theta_1}(j) U^+_{\Theta_2}(1-j) +
U^-_{\Theta_1}(j) U^-_{\Theta_2}(1-j) \right) \times \\
\nonumber & &\,\,
~~~~~~~
 \sin\phi_0\frac{q_c^{s^2/(k-2)}}{\vartheta_{11}(\frac{\phi_0}{\pi}|it_c)}.
\nonumber \eea The $x$-integral is performed in appendix A, \beq
\nonumber \mbox{($x$-integral)} =
\frac{\pi\beta}{|\sin\phi_0|}\delta(s)\,. \eeq Thus the overlap
in the winding number $w$ sector is given by \beq
\nonumber
 {}_{\Theta_1,w}\bra\bra B, \beta
|(q_c^{1\over2})^{L_0+\bar L_0 -{c\over12}} |B,
\beta\ket\ket_{\Theta_2,w}
 \propto   \beta \frac{ e^{-\frac{kt_c\beta^2}{8\pi}w^2}}{|\vartheta_{11}(\frac{\beta t_c
 w}{2\pi}|it_c)|}.
\eeq Altogether we have, \beq\label{thermal}
{}_{\Theta_1}\bra\bra B; \beta|(q_c^{1\over2})^{L_0+\bar L_0
-{c\over12}}|B;\beta\ket\ket_{\Theta_2}    \propto
\sum_{w=-\infty}^{\infty}  \beta \frac{
e^{-\frac{kt_c\beta^2}{8\pi}w^2}}{|\vartheta_{11}(\frac{\beta t_c
w}{2\pi}|it_c)|}. \eeq 
Note that the dependence on $\Theta_1$ and $\Theta_2$ has 
disappeared after the $j$-integral. In fact,
the expression (\ref{thermal}) 
agrees precisely with the result of \cite{lopt}
in the case of $\Theta_1 = \Theta_2=0$, $i.e.$ when 
the branes carry no fundamental string charge.  
There the annulus 
amplitude is evaluated exactly using the iterative Gaussian
integral method developed in \cite{Gawe,second}.
To compute the one loop free energy, we need to
multiply the partition functions of the ghost sector and the
internal CFT and to integrate the result over the moduli
space of the annulus worldsheet. The method has been explained
in detail in the appendix of \cite{lopt}, and we do not 
repeat it here.

We have found that the boundary state $|B; \beta\rangle\rangle_{\Theta}$
precisely reproduces the one loop free energy computed in
\cite{lopt} when $\Theta_1=\Theta_2=0$. 
It turned out that the partition function 
does not depend on $\Theta_1$ or $\Theta_2$ at all.
The reason for this is unclear and deserves a closer inspection.

\section{Conclusion}

In this paper, we constructed the boundary states for $AdS_2$ branes in
$AdS_3$, following \cite{chicagoone,chicagotwo} but using the different
ansatz. The boundary states are expressed as linear combinations
of the Ishibashi states as in (\ref{bs}), where the one point functions
$U^\pm_\Theta(j)$ are given by,
\beq \label{finalexpression}
  U_\Theta^\pm(j) 
= \Gamma\left( 1 - {2j-1 \over k-2}\right) \nu^{{1\over 2} - j}
e^{\pm \Theta(2j-1)},
\eeq
modulo factors independent of $j$. In the semi-classical approximation,
the location of the brane is given by $\psi = \Theta$ in the $AdS_2$
coordinates defined in Appendix B.  

From the point of view of the boundary conformal field theories,
the $AdS_2$ branes create defects which connect different conformal
field theories while preserving at least one Virasoro algebra. 
Since the boundary states allow study of the $AdS_2$ branes beyond
the supergravity approximation, it would be interesting to use them
to explore the correspondence further.  

\section*{Acknowledgments}

We would like to thank C. Bachas, J. de Boer, and R. Dijkgraaf 
for useful discussion on the holographic interpretation of
the $AdS_2$ branes. 
This research was supported in part by 
DOE grant DE-AC03-76SF00098. 

\newpage
\appendix
\section{Some useful integrals}
\subsection{Integral used in the semi-classical approximation}

We start with the following integral formula, 
 \beq
  \int_0^\infty dx \frac{x^n}{(ax^2+2bx+c)^{n+3/2}}={n!\over (2n+1)!!
  \sqrt{c} \left(\sqrt{ac}+b\right)^{n+1}}\,,
 \eeq
where $a\ge 0, c> 0, b> -\sqrt{ac}$. We want to integrate
 \beq
  \int_0^\infty dx_2^\prime \,\, \left(x_2'\right)^{2j-2}(x_2'^2-2\sinh\psi
  x_2'+\cosh^2\psi)^{{1\over2}-2j}\,.
 \eeq
Using $n=2j-2$, $a=1$, $c=\cosh^2\psi$ and $b=-\sinh\psi$, we see
the restrictions are clearly satisfied for all values of $\psi \in
{\mathbf R} $. Using the fact that $\Gamma\left(1/2+n\right)
= \sqrt{\pi} 2^{-n} (2n-1)!!$, we have
 \beq
   \int_0^\infty dx_2^\prime \, \left(x_2'\right)^{2j-2}(x_2'^2-2\sinh\psi
  x_2'+\cosh^2\psi)^{{1\over2}-2j}  ={2^{2-2j}\Gamma(2j-1)\sqrt{\pi}
   e^{\psi(2j-1)}\over \Gamma\left(2j-{1\over 2}\right)
   \cosh\psi}.
 \eeq

\subsection{Integral used for finite temperature free energy}
The integral we want to compute is: \beq \int_{Im x >0,\,
Im(xe^{i\phi_0})>0,\, e^{-\beta}<|x|<1} d^2 x |x-\bar
x|^{-2j}|xe^{i\phi_0}-\bar xe^{-i\phi_0}|^{-2(1-j)}. \eeq Use
polar coordinates such that $x=re^{i\theta}$.  Then we get, 
\bea &=&
\left( \int_{e^{-\beta}}^1 {dr \over r}\right) \int_0^{\pi-\phi_0}
d\theta |\sin\theta\sin(\theta+\phi_0)|^{-1}\exp\left(
2is\ln\left|\frac{\sin(\theta+\phi_0)}{\sin\theta}\right| \right)\nonumber \\
 &=& {\beta \over 2 \sin \phi_0}
\int_{-\infty}^\infty d\tilde s  e^{-is\tilde s}  
\frac{\pi}{|\sin\phi_0|} \\
&=&  \frac{\pi\beta}{|\sin\phi_0|} \delta(s) \, , \eea 
where, in the second line, we changed the variable 
$\theta$ to $\tilde s$ defined by
\beq \tilde 
s = 2\ln \left|{\sin (\theta+ \phi_0)\over \sin \theta}\right| .
\eeq

\section{Coordinate Systems for $AdS_3$}
\label{coordinates}

The space $AdS_3$ is defined as the
hyperboloid
\beq
\label{hyper}
(X^0)^2 - (X^1)^2 - (X^2)^2 + (X^3)^2 = R^2 \,,
\eeq
embedded in ${\bf R}^{2,2}$.  The metric
\beq
ds^2 = -(dX^0)^2 + (dX^1)^2 + (dX^2)^2 - (dX^3)^2
\eeq
on ${\bf R}^{2,2}$ induces a metric of constant negative curvature on
$AdS_3$. The quantity $R$ that appears in \eq{hyper} is the {\em
anti-de~Sitter radius}; for convenience, we set $R=1$.  In addition, to avoid
closed time-like curves, we work not with the hyperboloid
\eq{hyper} itself, but with its universal cover. 

The two coordinate systems we use most extensively are {\em global
coordinates} and {\em $AdS_2$ coordinates}.  The global coordinates
$(\r, \th, \t)$ are defined by
\beq
X^0 + i X^3 = \cosh \r \, e^{i t} \,, \qquad X^1 + i X^2 = -\sinh \r \,
e^{-i \th} \,.
\eeq
The range of the radial coordinate $\r$ is $0 \le \r < \infty$; the
angular coordinate $\th$ ranges over $0 \le \th < 2 \pi$; and the
global time coordinate $t$ may be any real number.
The $AdS_3$ metric in global coordinates is
\beq
ds^2 = - \cosh^2 \!\r \, dt^2 + d\r^2 + \sinh^2 \!\r \, d \th^2 \,.
\eeq

The 
$AdS_2$ coordinate system is convenient to use when
 considering $AdS_2$ branes
in $AdS_3$. They are defined by
\beq
X^1 = \cosh \psi \sinh \w \,, \qquad X^2 = \sinh \psi \,, \qquad X^0 +
i X^3 = \cosh \psi \cosh \w \, e^{it} \,.
\eeq
All three $AdS_2$ coordinates range over the entire real line. In this
parametrization, the fixed $\psi$ slices have the geometry of
$AdS_2$.  The $AdS_3$
metric in $AdS_2$ coordinates takes the form
\beq
ds^2 =  d\psi^2  + \cosh^2 \psi \, (-\cosh^2 \w \, dt^2 + d\w^2) \,;
\eeq
the quantity in parentheses is the metric of the $AdS_2$ subspace at
fixed $\psi$.   The transformation between global and $AdS_2$
coordinates is
\beq
\sinh \psi = \sin \th \sinh \r \,,\qquad \cosh \psi \sinh \w = - \cos
\th \sinh \r \,.
\eeq
The global time $t$ is the same in both coordinate systems.

The space $AdS_3$ is the group manifold of the group $SL(2,R)$.  A
point in $AdS_3$  is given by the $SL(2,R)$ matrix
\beq
g = \left( \begin{array}{cc} X^0 + X^1 & X^2 + X^3 \\ X^2 - X^3 & X^0
- X^1 \end{array} \right) \,.
\eeq
In the global coordinate system,
\beq
g =  \left( \begin{array}{cc} \cos t \cosh \r -
\cos \th \sinh \r & \sin t \cosh \r + \sin \th \sinh \r \\ -\sin t
\cosh \r + \sin \th \sinh \r & \cos t \cosh \r + \cos \th \sinh \r
 \end{array} \right) \,.
\eeq

\medskip

In this paper,
we work mostly in the Euclidean rotation of this geometry,
which is given by $t \rightarrow t_E = it$ in the above.

\section{Some useful formulae}
\label{hypergeometric}
In this appendix, we list some useful formulae involving
the hypergeometric function and the gamma function.

The hypergeometric function, ${}_2F_1(\a,\b,\g;z)$, enjoys 
following useful identities:
\bea
   {}_2F_1(\a,\b,\g;z)=(1-z)^{\g-\a-\b} {}_2F_1(\g-\a,\g-\b,\g;z).
\eea
The Gauss recursion formulae are given by
\bea
  && \g\, {}_2F_1(\a,\b,\g;z)+(\b-\g)\,{}_2F_1(\a+1,\b,\g+1;z)-
   \\ \nonumber 
  && \qquad \qquad \qquad \qquad \b(1-z)\, {}_2F_1(\a+1,\b+1,\g+1;z)=0,  \\
 && \g\, {}_2F_1(\a,\b,\g;z)-\g\, {}_2F_1(\a+1,\b,\g;z) +\b z
   \, {}_2 F_1(\a+1,\b+1,\g+1;z)=0, \\
 && \g\, {}_2F_1(\a,\b,\g;z)-(\g-\b)\,{}_2F_1(\a,\b,\g+1;z)-
   \b\,{}_2F_1(\a,\b+1,\g+1;z)=0.
\eea
Under $z\rightarrow 1-z$, we have
\bea
  && {}_2F_1(\a,\b,\g;1-z)={\Gamma(\g) \Gamma(\g-\a-\b) \over
    \Gamma(\g-\a) \Gamma(\g-\b)} {}_2F_1(\a,\b,1-\g+\a+\b;z) + \\
  \nonumber
  &&  \qquad \qquad z^{\g-\a-\b} {\Gamma(\g) \Gamma(\a+\b-\g) \over \Gamma(\a)
    \Gamma(\b)} {}_2F_1(\g-\a,\g-\b,1+\g-\a-\b;z).
\eea
Lastly, under $z \rightarrow 1/z$, we get
\bea
  && {}_2F_1(\a,\b,\g;z)={\Gamma(\g) \Gamma(\b-\a) \over
    \Gamma(\b) \Gamma(\g-\a)} \left(-{1\over z}\right)^{\a} 
    {}_2F_1 \left(\a,1+\a-\g,1+\a-\b;
    {1\over z}\right) + \\ \nonumber
  && \quad \qquad {\Gamma(\g) \Gamma(\a-\b) \over \Gamma(\a) \Gamma(\g-\b)}
    \left(-{1\over z}\right)^\b 
    {}_2F_1 \left(\b,1+\b-\g,1+\b-\a;{1\over z}\right).
\eea
We list some useful identities involving the gamma, $\Gamma(z)$, function:
\bea
 && \Gamma(1+z)=z \Gamma(z), \\
 && \Gamma\left({1\over 2}\right) = \sqrt{\pi}, \\
 && \Gamma(1-z) \Gamma(z) = {\pi \over \sin(\pi z)}, \\
 && \Gamma(1+ix) \Gamma(1-ix) = {\pi x \over \sinh(\pi x)}, \qquad x\in {\bf R} \\
 && \Gamma(2z) = {2^{2z-1}\over \ \sqrt{\pi}} \Gamma(z) \Gamma
 \left(z+{1\over 2}\right).
\eea
\noindent
\begingroup\raggedright\endgroup

\end{document}